\newcommand{\OIII}{[OIII]~$\lambda 5007\,{\rm\AA}$~}
\newcommand{\ltsima}{$\; \buildrel < \over \sim \;$}
\newcommand{\simlt}{\lower.5ex\hbox{\ltsima}}            % < over ~
\newcommand{\gtsima}{$\; \buildrel > \over \sim \;$}
\newcommand{\simgt}{\lower.5ex\hbox{\gtsima}}            % > over ~
\newcommand{\arcmin}{\rm arcmin}
\documentstyle[graphics]{mn}
\title[Intergalactic stars in Fornax]
{Intergalactic Stars in the Fornax Cluster
\thanks{Based on observations collected at the European Southern Observatory, 
La Silla, Chile}}

\author[T. Theuns and S. J. Warren]
{Tom Theuns$^{\,1,2}$ and S. J. Warren$^{\,3}$\\
$^1\,$ University of Oxford, Department of Physics, Astrophysics, Keble
Road, Oxford OX1 3RH
\\
$^2\,$ Scuola Normale Superiore di Pisa, Piazza dei Cavalieri 7,
I-56126 Pisa, Italy\\
$^3\,$ Blackett Laboratory, Imperial College of Science Technology and Medicine,
Prince Consort Road, London SW7 2BZ}

\begin{document}

\maketitle

\begin{abstract}
We have identified ten candidate intergalactic planetary nebulae in
the Fornax galaxy cluster. These objects were found during
observations in 1992 and 1993 in three fields chosen well away from
any Fornax galaxy at 15\arcmin, 30\arcmin, and 45\arcmin~ from the
centre of Fornax. We used the usual method of blinking images taken in
a narrow \OIII filter, with images taken in an adjacent broad
filter. The measured fluxes in the narrow, broad, and I bands are
consistent with these unresolved objects being planetary nebulae
immersed in an intergalactic population of stars. Such a population is
expected to arise as a consequence of tidal encounters between
galaxies, and our findings strengthen the case for the existence of
such tidal debris. The confirmation of some or all of these ten
candidates as planetary nebulae would imply that intergalactic stars
constitute a substantial fraction of all the stars in Fornax, up to an
estimated $\sim40$ per cent. Intergalactic planetary nebulae could prove
useful in probing the underlying cluster potential, since they would
be far more abundant than galaxies. We discuss possible contamination
of the sample by emission-line galaxies, but conclude that planetary
nebulae is the most likely identification for the detected objects.
\end{abstract}


\section{Introduction}
Tidal interactions between galaxies in a cluster are expected to be
frequent and may induce structural changes in the disturbed galaxies.
For example, numerical simulations by Moore et al. (1996) indicate that
fast tidal encounters produce \lq harassed\rq~ looking galaxies,
reminiscent of the distorted spirals in clusters at a redshift of
$\sim$ 0.4 as observed by the Hubble Space Telescope. Numerical
simulations of the evolution of galaxies in a cluster (e.g. Roos \&
Norman 1979, see Dressler 1984 for a review) suggest that these tidal
encounters cause galaxies to loose a substantial fraction, $30-70$ per
cent, of their stars to the cluster potential where they are left free
to roam as intergalactic tidal debris.

Zwicky (1951) associated excess light he had observed between the
galaxies in Coma with such debris from tidal interactions. Subsequent
searches for intergalactic light in clusters of galaxies (usually in
Coma) and Hickson's compact groups, using either photographic plates,
direct photometric observations or CCD imaging, were not all equally
conclusive (e.g. de Vaucouleurs \& de Vaucouleurs 1970, Vilchez, Pello
\& Sanahuja 1994, and references therein), with many observers
concluding that the amount of intergalactic light could be attributed
to low surface brightness galaxies, inaccurate subtraction of
foreground stars and halos of galaxies. Another line of attack to track
down the debris is to look for intergalactic supernovae. Of the
thirteen supernovae detected in Coma over a period of fifteen effective
search years, Crane, Tammann and Woltjer (1977) could associate all of
them with galaxies, either in Coma or in the background. They concluded
that the intergalactic debris in Coma is at least six times less active
per unit light in producing supernovae than the Coma galaxies
are. Harris (1986), following up ideas of Merritt (1983, 1984),
suggested that the central galaxy in a cluster might be surrounded by a
population of globular clusters, stripped from their parent
galaxies. This proposal was investigated by Muzzio (1987, and
references therein) and more recently by West et al. (1995), who
suggest that the pronounced globular cluster overabundance of some
galaxies in the centres of galaxy clusters is best explained by
assuming the existence of a population of intergalactic globular
clusters.

In this Letter we report on the identification of ten candidate
intergalactic planetary nebulae (PNe) in the Fornax galaxy
cluster. Fornax ($\alpha$=$3^h 38^m 29^s$ $\delta$=$-35^\circ 27^{'}$
(J2000.0), distance $17\pm 1$Mpc, McMillan, Ciardullo and Jacoby 1993)
is a poor, dense cluster of galaxies, relatively rich in early type galaxies
(ratio of early to late types a factor of two higher than in Virgo,
Ferguson 1989). It has a high central density of galaxies (500 per
Mpc$^3$) and a low velocity dispersion (350 ${\rm km
\,s^{-1}}$, Ferguson 1989). Its central galaxy, NGC 1399, is a cD
galaxy with both an extended halo of diffuse light (Schombert 1986,
Warren, Theuns \& Williger 1996) and an abundant and extended globular
cluster population (Grillmair et al. 1994 and references therein).
All these properties suggest that Fornax should have a substantial
population of intergalactic stars. Since these stars should be similar
to the stars in the galaxies, they would also have a sub-population of
PNe. The detection of intergalactic PNe as far away as Fornax is
feasible technically, since PNe within galaxies in Fornax have already
been found. McMillan et al. (1993) used narrow-band imaging to
identify 72 candidate PNe in NGC 1399. Arnaboldi et al. (1994) later
obtained spectra for 54 of them and they were able to obtain
velocities for a subset of 37. We have used the same technique to
search for intergalactic stars.

In Section~2 of this Letter we describe our observational strategy to
detect the brightest PNe in three fields in the Fornax cluster, and detail the
standard reduction and selection procedures. Section~3 summarizes our
findings, and Section~4 contains a discussion of our results.

\begin{figure*}
\setcounter{figure}{0}
\setlength{\unitlength}{1cm}
\centering
\begin{picture}(16,21)
\put(0.0,2.0){\includegraphics{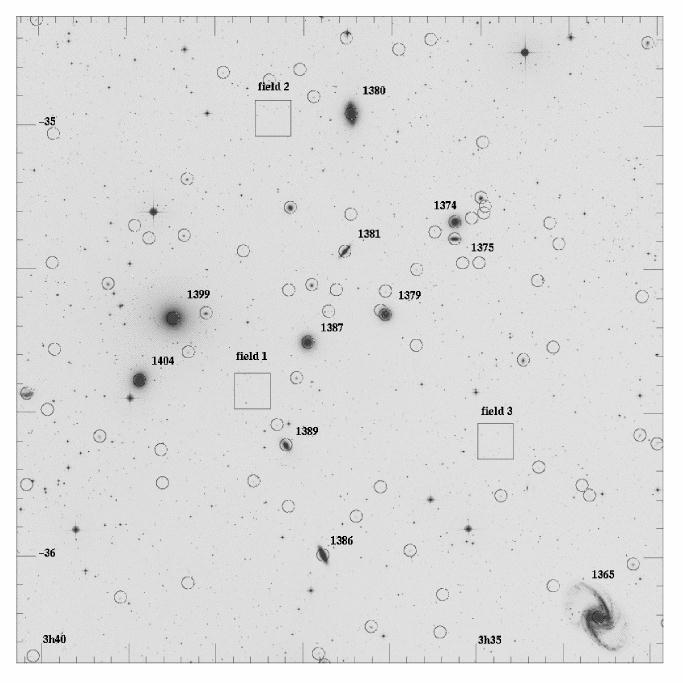}}
%\put(0.0,2.0){\special{psfile=thumb.ps vscale=80 hscale=80
%angle=0 voffset=-160 hoffset=0}}
\end{picture}
\caption{A region of size $90$\,arcmin in Fornax
drawn from the Digitized Sky Survey (J2000.0 equinox). The major Fornax
galaxies are indicated by their NGC numbers; NGC~1399 is usually taken
to be the centre of Fornax. Circles denote galaxies which are likely to
be Fornax members, according to Ferguson (1989). The three fields are
indicated by a square of side 5.9\arcmin. For the adopted distance
$d=17$~Mpc, the field of view corresponds to 445~kpc on a side.}
\label{fig:fornax}
\end{figure*}

\section{Observations}
\subsection{Set-up and data reduction}
We have imaged three fields in the Fornax cluster in narrow, broad, and
I bands, using the EMMI instrument on the New Technology Telescope at
the European Southern Observatory, La Silla, Chile. The fields are
located at angular separations of 15\arcmin~, 30\arcmin~, and
45\arcmin~ from the centre of NGC~1399. The coordinates of the fields
are provided in Table~\ref{table:radec}, and the locations are shown
in Fig.~\ref{fig:fornax}. The dates of the observations were 1992
December 30 and 31, and 1993 November 19, 20, and 21. Conditions were
mostly clear, with some thin cirrus at times.  Details of the average
seeing (image FWHM in the combined frames), and integration times are
provided in Table~\ref{table:log}. The narrow-band and broad-band
filters were used in the blue arm (BIMG), where the scale is
0.37\arcsec\, per pixel. The I-band observations were made with the
red arm (RILD), with the old $f/2.5$ camera, for which the scale was
0.44\arcsec\, per pixel in 1992, with the Thomson chip, and
0.35\arcsec\, per pixel in 1993, with the Ford chip.

\begin{table}
\centering
  \caption{Coordinates of observed fields\label{table:radec}}
   \begin{tabular} {@{}ccc} 
\multicolumn{1}{c}{field}&\multicolumn{1}{c}{RA}&\multicolumn{1}{c}{Dec.}
\\
\multicolumn{1}{c}{ }&\multicolumn{2}{c}{J2000.0 equinox} \\ \hline
1   & 3 37 35 & $-35$ 37 15 \\
2   & 3 37 17 & $-34$ 59 10 \\
3   & 3 34 48 & $-35$ 44 05 \\
   \end{tabular}
\end{table}

The narrow-band filter has FWHM $29{\rm\AA}$ and is centred at
$\lambda_c=5028{\rm\AA}$, very close to the wavelength of \OIII
redshifted to the velocity 1422 ${\rm km \,s^{-1}}$ of NGC~1399. In
EMMI the blue filters lie in a diverging $f/11$ beam, so the
broadening and blue shift of the transmission curve, relative to the
laboratory curve (measured in a parallel beam), may be neglected. The
broad-band filter has FWHM $286{\rm\AA}$ and is centred at
$\lambda_c=4714{\rm\AA}$.  Integration times, filter transmission,
telescope size, and seeing are comparable to those for the runs of
McMillan et al. (1993), so we expect our observations to reach
slightly fainter than theirs since they had to find PNe on top of the
bright parent galaxy.

\begin{table}
 \begin{minipage}{80mm} 
  \caption{Observing log\label{table:log}}
   \begin{tabular} {@{}cccccr} 
\multicolumn{1}{c}{field}&\multicolumn{1}{c}{date}&\multicolumn{1}{c}{filter}&
\multicolumn{1}{c}{no. of}&\multicolumn{1}{c}{average}&\multicolumn{1}{c}
{total}\\
\multicolumn{1}{c}{ }&\multicolumn{1}{c}{ }&\multicolumn{1}{c}{ }&
\multicolumn{1}{c}{frames}&\multicolumn{1}{c}{seeing}&\multicolumn{1}{c}
{int. time}\\
\multicolumn{1}{c}{ }&\multicolumn{1}{c}{ }&\multicolumn{1}{c}{ }&
\multicolumn{1}{c}{ }&\multicolumn{1}{c}{arc sec.}&\multicolumn{1}{c}
{s}\\ \hline
1   & Nov 93  &   narrow  & 3  &  1.0   & 9900\\
1   & Nov 93  &   broad   & 4  &  1.0   & 7200\\
1   & Nov 93  &   I       & 2  &  1.1   & 720 \\
2   & Dec 92  &   narrow  & 4  &  1.2   & 13200 \\
2   & Dec 92  &   broad   & 5  &  1.3   & 6300\\
2   & Dec 92  &   I       & 4  &  0.9   & 960 \\
3   & Nov 93  &   narrow  & 4  &  1.0   & 13200 \\
3   & Dec 92  &   broad   & 2  &  1.1   & 2400\\
3   & Nov 93  &   broad   & 5  &  1.1   & 6000\\
3   & Dec 92  &   I       & 4  &  0.9   & 960 \\
   \end{tabular}
 \end{minipage}
\end{table}

After the usual dark and bias subtraction and flat fielding, we
combined the individual images for a given field.  The combining
procedure weights the frames by the inverse of the variance in the
sky, and also compares the respective pixel values for the individual
frames to detect and eliminate cosmic rays. The final frames in each
field cover an area $5.9^{\prime}\times5.9^{\prime}$.  We flux
calibrated the broad-band and narrow-band frames onto the AB system
using observations of the standard stars Hiltner 600 and LTT1788
(Hamuy et al., 1992). For the I-band frames we used standards from
Landolt (1992), on the Kron-Cousins system.  Conditions were not
always photometric. However, because we have a number of frames in
each field, as well as multiple obervations of standard stars, we were
able to keep track of the extinction due to clouds, which was only
ever at the level of 0.1 mag or less. By identifying those
observations taken when the sky was clear we were able to calibrate
the fields, and we estimate that the photometric zero points are
accurate to 0.05 mag. This is borne out by a comparison between the
fields of the median narrow-broad colour $m_n-m_b$ of all the objects
in the field. The scatter between the three fields of the median
colour is $\sigma=0.04$ mag.

\subsection{Analysis}

% here come all the cm and cc figures
\begin{figure*}
\setcounter{figure}{1}
\setlength{\unitlength}{1cm}
\centering
\begin{picture}(15,8)
\put(-1.5,-3.5){\includegraphics{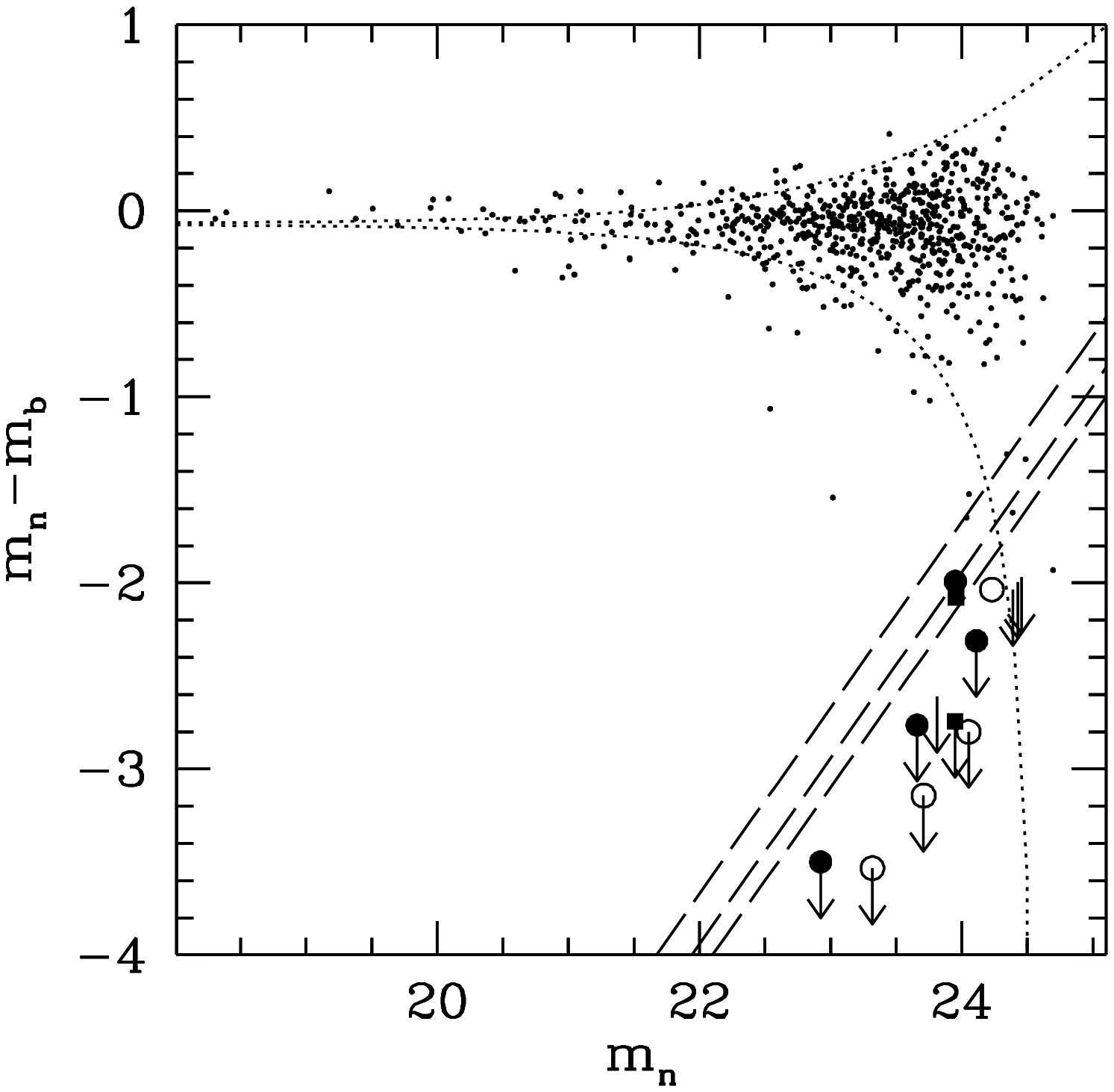}}
\put( 5.5,-3.5){\includegraphics{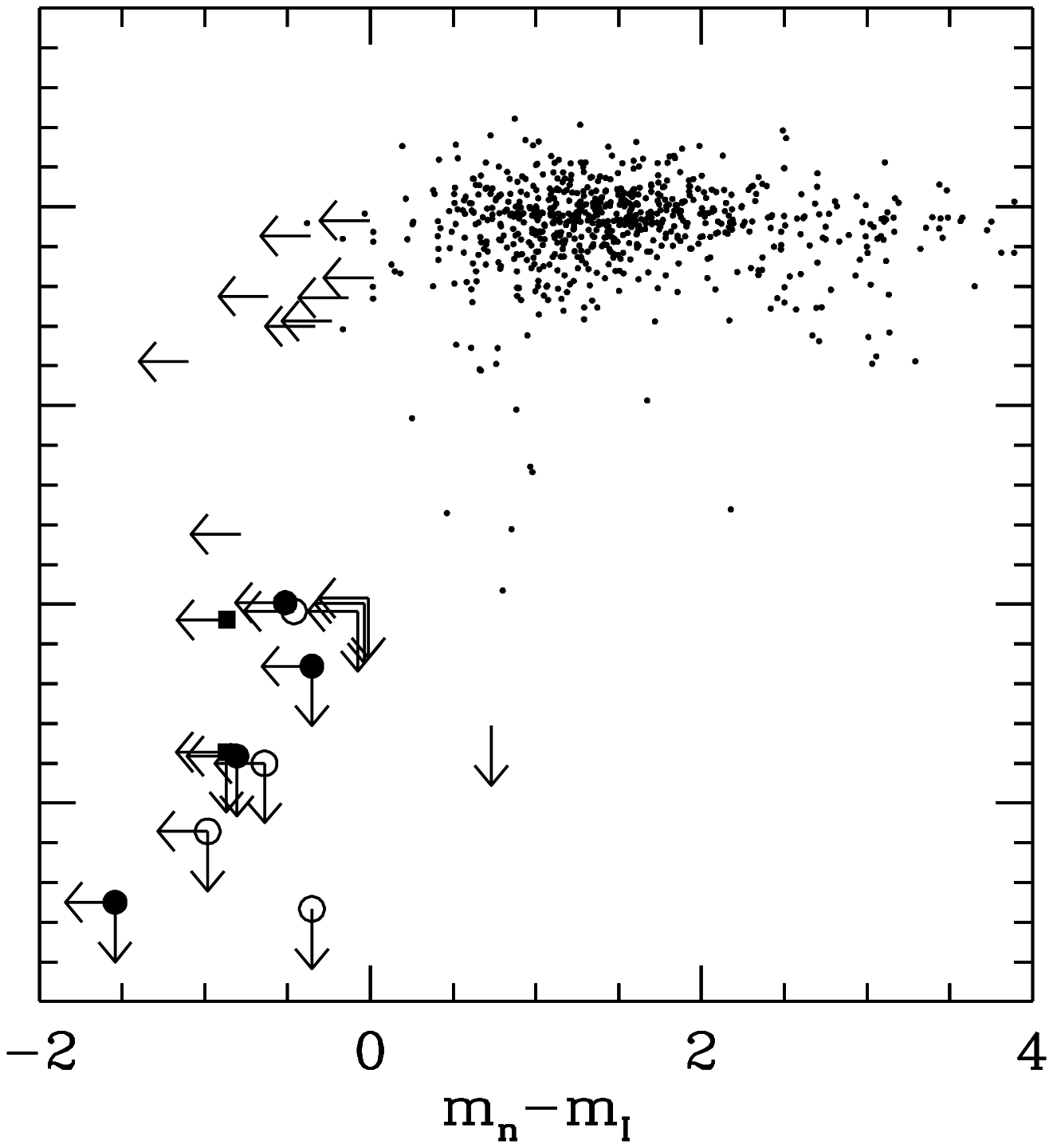}}
\end{picture}
\caption{Colour-magnitude (left panel) and colour-colour (right panel)
diagrams for objects in fields 1--3. Dots denote objects detected in the
corresponding bands, arrows denote $1\sigma$ upper limits for
non-detections. Objects satisfying all selection criteria (see text)
are plotted as filled circles (field 1), filled squares (field 2) and
open circles (field 3). Left panel: objects within  $3\sigma$ of
the median colour fall inside dotted lines. The dashed lines show the
$2\sigma$ detection limits for the broad band, for each
field. Candidate intergalactic PNe are objects that fall below both
these selection limits, and are bluer than $m_n-m_I=0.0$.}
\label{fig:colmag}
\end{figure*}

From a knowledge of the luminosity function of PNe, the exposure times,
the distance to Fornax, and the very large expected equivalent widths
(EWs) of the \OIII line, most likely any PNe visible in our narrow-band
frames will be absent from the broad-band and I-band frames. Based on
this, we have used the following procedure to identify candidate
PNe. Using the photometry package Daophot (Stetson 1987), we produced a
source list for each narrow-band frame, restricted to objects detected
at greater than $4\sigma$ significance, as measured by aperture
photometry using an aperture of diameter twice the stellar FWHM. Next
resolved sources were eliminated, using the Daophot procedure {\it
peak,} which compares the profile of each object against the
point-spread-function obtained from bright stars. Using a six-parameter
transformation the coordinates of these sources in the other bands were
computed, and the magnitudes were measured at these positions,
i.e. without recentering, again using an aperture of twice the stellar
FWHM for the particular frame. This procedure is designed for sources
that are very faint in these bands, for which the centering algorithm
would not work well. Finally, we assigned fluxes corresponding to
1~$\sigma$ above sky to objects with measured fluxes below this
level. This produced magnitudes in the narrow, broad, and I bands
(denoted here by $m_n$, $m_b$, $m_I$ respectively) for our basis set of
objects. The $1\sigma$ detection levels, averaged for the three fields,
are $m_n=26.0, m_b=26.7, m_I=24.7$.

\section{Results}

The results of the photometry are illustrated in
Fig.~\ref{fig:colmag}, which shows $m_n-m_b\, v\, m_n$
colour-magnitude and $m_n-m_b\, v\, m_n-m_I$ colour-colour diagrams.
These diagrams can be used to select candidate intergalactic PNe. Any
PNe in our frames will have unusual colours relative to the locus of
common objects visible in the plots, which comprise Galactic stars and
compact galaxies.  Candidates were selected using the following
criteria:

1.) the $m_n-m_b$ colour differs, in the negative direction, by more
than 3~$\sigma$ from the median $m_n-m_b$ for all objects in the frame
$-$ such objects fall below the lower dotted line in
Fig.~\ref{fig:colmag}, left panel.

2.) the objects are not detected at $\ge 2\sigma$ in the
broad-band frame $-$ such objects
fall below the dashed lines in Fig.~\ref{fig:colmag}, left panel.

3.) objects have $m_n-m_I<0.0$.

Criterion 1 requires that the candidates have significant excess flux in
the narrow-band frame relative to the flux in the broad-band
frame. Criterion 2 states that the line EWs are consistent with being
large: at the $2\sigma$ level they are consistent with being infinite.
Criterion 3 is an attempt to remove compact emission-line galaxies
from the sample.  In particular galaxies in these fields at $z=0.35$,
for which the line [OII] $\lambda 3727{\rm\AA}$ falls in the
narrow-band filter, but which have a substantial $4000{\rm \AA}$
break, will be eliminated by this means.

Application of the above selection criteria leads to the
indentification of 4, 2, and 4 candidates in fields 1, 2, and 3
respectively.  We have checked the images of each candidate in every
CCD frame used to make the combined frames, to ensure that the unusual
colours are not due to flaws in individual frames, or e.g. cosmic
rays, or satellite trails. The candidates have narrow-band magnitudes
in the range $22.9<m_n<24.2$. The estimated line fluxes \footnote{The
line fluxes were calculated from $F_{5007}=f_{\lambda}\int
T_{\lambda}d\lambda/T_L$, where $f_{\lambda}$ is the flux per unit
wavelength of a continuum source with the same $m_n$, $T_{\lambda}$ is
the transmission of the filter as a function of wavelength, and $T_L$
is the average transmission over the central few angstroms.} are in
the range $9.5\times 10^{-17}>F_{5007}>2.9\times 10^{-17}$ erg
cm$^{-2}$ s$^{-1}$, and $26.3<m_{5007}<27.6$, where as usual
$m_{5007}=-13.74-2.5{\rm log}F_{5007}$. None of the candidates is
detected at greater than $2\sigma$ significance in the broad-band. In
the I band the brightest candidate in field~3 is possibly detected at
$2.5\sigma$, while the remaining 9 candidates are fainter than the
$1\sigma$ level.

\section{Discussion}

To summarise the previous sections, we have discovered 10 objects in
our fields whose angular sizes and magnitudes in the three bands are
consistent with their being intergalactic PNe in the Fornax
cluster. We now consider in more detail whether any of the candidates
could be distant galaxies, rather than PNe, because either [OII] 3727
($z=0.35$) or Ly$\alpha$ 1216 ($z=3.14$) falls within the narrow-band
filter. The question is not straightforward to answer as there are few
surveys which reach the faint line-flux limits of our narrow-band
images, and the continuum fluxes of our candidates, $m_b>26$,
$m_I>24$, are beyond the limits of the faintest magnitude-limited
spectroscopic surveys. Nevertheless the lower redshift appears
unlikely. The mean rest-frame EW of the candidates if at $z=0.35,$ is
$>250{\rm\AA}$, and such large EWs have not been seen for [OII] in
deep galaxy surveys (Colless et al. 1990, Cowie et al. 1995).
Furthermore galaxies at this redshift would likely be resolved in our
frames.

We can check whether the higher redshift $z=3.14$ is a possibility by
making a direct comparison against the results of the emission-line
searches for high-redshift galaxies by Lowenthal et al. (1995) and
Thompson et al. (1995). On the basis of their quoted detection limits,
if our candidates are galaxies at $z=3.14$, we estimate that in
each of these surveys three galaxies $z>2$ would have been found above
the 3$\sigma$ detection limits, whereas none was discovered. This is
fairly clear evidence against our candidates being galaxies at
$z=3.14$.

On the basis of the foregoing we consider it likely that our
candidates are intergalactic PNe. We have also compared the
distribution in luminosity of our candidates against the luminosity
function of Ciardullo et al. (1989). A K-S test indicates a
satisfactory fit, which increases our confidence in our
identification. Just before completing this Letter we received a
preprint by Arnaboldi et al. (1996) who, while searching for PNe in
the outer halo of the Virgo elliptical NGC~4406, serendipitously found
three PNe with radial velocities widely different from that of the
target galaxy. They concluded that these PNe may be intergalactic
stars in the Virgo cluster. With the wide-field CCD cameras becoming
available it appears that it may now be possible to detect
intergalactic PNe in large numbers.

Proceeding on the assumption that our 10 emission-line candidates are
intergalactic PNe, we next try to estimate the contribution of
intergalactic stars to the total light in the Fornax cluster. Our
observations reach about one magnitude below $M^*$ in the PNe
luminosity function, and we adopt the value for the luminosity-specific
planetary nebula density for M31, $\alpha_{1.0(B)}=9.4\,\times
10^{-9}/(L_B)_\odot$ (Ciardullo et al. 1989). (We note, however, that the
value of $\alpha_{1.0(B)}$ is probably different for different stellar
populations (Hui et al. 1993), and therefore the chosen value may not be
appropriate for intergalactic debris.) We surveyed 104 arcmin$^2$, so
this provides an estimate for the surface brightness in tidal debris
of $1.0\,\times 10^{7}(L_B)_\odot$ per arcmin$^2$. The total
luminosity contributed by galaxies contained in the central 200\arcmin
$\times$ 200\arcmin\, region of Fornax is $\approx 6.6\times 10^{11}
(L_B)_\odot$ (Caldwell 1987), corresponding to a surface brightness of
$\sim 1.65\times 10^7\,(L_B)_\odot$ per arcmin$^2$. Comparing these
two values we estimate that the contribution of intergalactic stars to
the total light in the Fornax cluster could be as much as $\sim 40\%$.

To summarise, with narrow-band imaging we have searched for
intergalactic PNe in the Fornax cluster, and discovered 10 good
candidates, which require spectroscopic confirmation. Future surveys
could detect hundreds of such objects. Intergalactic PNe may,
therefore, prove invaluable in the study of cluster dynamics.

\section*{Acknowledgments}
During the course of this work, TT was funded by a studentship at ESO
and a PPARC post-doc at Oxford.  TT is presently funded by the EC
under contract CT941463. We thank the referee, Russet McMillan, for a
number of suggestions which improved the presentation of this
Letter. We are grateful to Bo Reipurth, who agreed to swap
observing time in 1992, when late delivery of our filters made it
impossible to start observing. TT wants to thank Mike Barlow, Robin
Clegg and Mario Livio for valuable discussions. Fig.~\ref{fig:fornax}
was extracted from the Digitized Sky Survey, which was produced at the
Space Telescope Science Institute under U.S. Government grant NAG
W-2166.

{}

\end{document}